# Mitigation of calibration ringing in the context of the MTG-S IRS Instrument

**Pierre Dussarrat [1*], Guillaume Deschamps [1], Bertrand Theodore [1], Dorothee Coppens [1], Carsten Standfuss [2], and Bernard Tournier [2]**

[1]EUMETSAT, Eumetsat-Allee 1, 64295 Darmstadt, Germany
[2]SPASCIA, 14 Avenue de l'Europe, 31520 Ramonville-Saint-Agne, France
*Correspondence: pierre.dussarrat@eumetsat.int

**Abstract:** EUMETSAT is currently developing the on-ground processing chain of the infrared Fourier transform spectrometers (IRS) on-board of the Meteosat Third Generation sounding satellites (MTG-S). In this context, the authors have investigated the impact of a particular type of radiometric error, called hereafter the calibration ringing. It arises in Fourier transform spectrometers when the instrument Radiometric Transfer Function (RTF) varies within the domain of the instrument Spectral Response Function (SRF). The expected radiometric errors are simulated in the context of the MTG-S IRS instrument in the long wave infrared (LWIR) band. Making use of a principal components (PC) decomposition, a software correction, called RTF uniformisation, is designed and its performance is assessed in the context of MTG-S IRS.

## 1. Introduction

Fourier transform infrared spectrometry from space allows decomposing the light exiting the atmosphere from its top: the reconstructed spectra exhibit absorption and emission lines representative of the Earth's atmosphere composition and thermodynamic state. Instruments, such as the Infrared Atmospheric Sounding Interferometer (IASI) on Metop [1] and the Cross-track Infrared Sounder (CrIS) on Suomi-NPP [2], have proven very effective for both weather forecasting and climate monitoring. EUMETSAT is now preparing for the next generation of instruments and in particular, for the first European Fourier transform spectrometer on a geostationary orbit, the infrared Fourier transform spectrometers (IRS) on-board of the Meteosat Third Generation sounding satellites (MTG-S) [3]. The ambition to generate spectra with a radiometric accuracy below the deci-Kelvin limit has brought attention to a subtle radiometric bias in Fourier transform spectrometry that we referred to hereafter as calibration ringing [4,5]. Such effect arises when the instrument radiometric transfer function (RTF) varies significantly within the domain of the instrument spectral response function (SRF). These variations generate distortions of the SRF [4], which, if unaccounted, propagate as radiometric errors into the calibrated Earth view spectra exploited by the users.

RTF spectral variations are actually expected for most instruments. As a first example, transmission cut-offs of the optical elements could lead to strong transmission gradients at the band edges and would produce calibration ringing. Thus, Borg et al [6] have recently reported the presence of such ringing in the spectra measured with the CrIS instrument that are induced by a steep instrument RTF at the beginning of the LWIR band. As a second example, RTF modulations can arise from unexpected light loop between optical surfaces. Optical elements in transmission such as lenses, windows or protective layers with non-perfect coating can create low finesse etalons [7]; as a result the transmission appears modulated as function of the incident light wavenumber. The latter effect is expected in the MTG-S IRS instrument, with a spectral modulation in wavenumber at approximately 4 mm frequency and up to a few percent relative amplitude. The optimization of the optical design of the instrument (e.g. re-designing the optical coating such that the RTF is flat) proved technically out of reach. Therefore, assuming a frozen design of the IRS instrument, the authors have initiated a study aiming at developing a methodology to mitigate the impact of calibration ringing and introduce the correction into the operational on-ground processing.

The fundamentals of calibration ringing are recalled in section 2. The discussion and definition of the mitigation strategy, called RTF uniformisation, follows in section 3. Finally, the last section presents simulations testing the RTF uniformisation in the context of MTG-S IRS long-wave infrared band (LWIR).

## 2. Calibration Ringing

Calibration ringing errors occur when the radiometric calibration fails to perfectly compensate for the RTF spectral variations. Usually, the instrument optical transmission is characterized in flight with specific calibration schemes using, for example, on-board black-body and deep space measurements [8]. The optical transmission is then removed from the raw Earth view measurements by division by the radiometric calibration factors. However, we still expect the occurrence of calibration ringing as high-frequency residual spectral modulations. In the following, we propose a simplified description of the radiometric calibration step of FTS products to highlight the genesis of calibration ringing.

The instrument SRF, noted SRF($\nu$) as function of the wavenumber $\nu$, is given in the context of Fourier transform spectrometry by the Fourier transform of the numerical apodisation $Apod(x)$ applied to the recorded interferograms as function of the Optical Path Distance $x$ (OPD). We assume hereafter for simplicity that self-apodisation defects are negligible or perfectly compensated such that the SRF is the same for all spectral channels.

The radiometric calibration factors are usually computed from the ratio between a spectrum measured from a black-body view and the associated theoretical Planck radiance at the blackbody temperature $\Gamma_{TBB}(\nu)$. As the blackbody radiation is flat at the scale of the SRF, the radiometric calibration factors $R(\nu)$ can be written as the RTF convoluted with the SRF only:

$$R(\nu) = \frac{[\Gamma_{TBB}.T \otimes SRF](\nu)}{\Gamma_{TBB}(\nu)} \cong [T \otimes SRF](\nu) \quad (1)$$

where $T(\nu)$ is the RTF of the instrument. If $Sp(\nu)$ is the radiance spectrum exiting the atmosphere as a function of the wavenumber, then the raw spectra are given by $[Sp.T \otimes SRF](\nu)$ and the radiometrically calibrated radiances $Sp_r(\nu)$ are written as:

$$Sp_r(\nu) \cong \frac{[Sp.T \otimes SRF](\nu)}{[T \otimes SRF](\nu)} \neq [Sp \otimes SRF](\nu) \quad (2)$$

Thus, if the optical transmission varies significantly at the scale of the SRF then the calibrated spectrum does not equal the input spectrum convoluted with the SRF as expected by the data users. The difference between the two terms defines the so-called calibration ringing error [4]. As illustrated in this paper, the calibration ringing radiometric errors usually appear as both a modulation and spikes that are functions of the input scene. Therefore, it cannot be canceled completely by a simple bias correction of the calibrated spectra and may affect the retrieval of the atmosphere composition.

As discussed in [6], considering the effect of the calibration ringing error in data applications is theoretically possible, for example by introducing the instrument transmission into the radiative transfer models (RTM) as in Equation (2). However, if the transmission varies between detectors, the RTM would become detector-dependent. In the case of the CrIS instrument, Borg et al [6] have demonstrated that, fortunately, one transmission would be sufficient to take into account the calibration ringing of the instrument to the nine instrument pixels. Nonetheless, in spectro-imagers such as MTG-S IRS in which a single acquisition consists of 25600 pixels, the etalon properties could strongly depend on the field of view; therefore, one transmission per pixel should be introduced into the RTM which proves computationally heavy and unpractical for the data users.

Consequently, in preparation for the on-ground processing of the MTG-S IRS data and expecting a possible strong impact of calibration ringing specifically on the LWIR products,

EUMETSAT has developed a mitigation algorithm for IRS LWIR band. Nonetheless, the methodology can be extended to its MWIR band and to other hyperspectral instruments.

## 3. RTF Uniformisation

As discussed in the introduction, the main contributor to calibration ringing for the CrIS instrument is the band cut-off close to the lower user band limit [6]. For MTG-S IRS, the main contributor is expected to be the etalon, the band cut-off effect being present but not dominating [4]. Therefore, this paper focuses primarily on RTF modulation induced by etalon effect. We first show with a basic example that additional spectral information is required to be able to correct the calibration ringing in presence of RTF modulation, then, we discuss a statistical method to estimate the (non-measured) high frequencies of a spectrum. Finally, we devise a correction factor to be applied to the calibrated measurement to mitigate the calibration ringing.

3.1. First insight

To gain insight on the impact of calibration ringing and its possible mitigation, we first consider a simple case for which the instrument transmission is modulated at a frequency $f$ and relative amplitude $\alpha$:

$$T(\nu) = 1 + \alpha \cos[2\pi\nu f] \quad (3)$$

The interferogram associated to the raw measurement in presence of RTF modulation $I(x)$ and the one associated to the measurement without RTF modulation $I_0(x)$ are given by Fourier transform of the spectra $Sp(\nu)$. They are representatives of the frequency decomposition of the spectra:

$$\begin{aligned} I(x) &= FT[Sp.T](x) \\ I_0(x) &= FT[Sp](x) \end{aligned} \quad (4)$$

Injecting the modulated transmission (Eq. 3), we get:

$$I(x) = I_0(x) + \frac{\alpha}{2} \times [I_0(x+f) + I_0(x-f)] \quad (5)$$

Thus, the interferogram recorded in presence of a RTF modulation is given by the superimposition of the one without modulation with two additional low amplitude interferograms shifted by the modulation frequency, called ghosts (Fig. 1).

Equation (5) can theoretically be reversed to retrieve $I_0(x)$ and thus the spectrum without ringing. At first order in $\alpha$, the approximate correction would write:

$$[Sp \otimes SRF](\nu) \cong FT^{-1}\left[\left\{I(x) - \frac{\alpha}{2} \times [I(x+f) + I(x-f)]\right\} \times Apod(x)\right] \quad (6)$$

Nonetheless, the samples above the maximum OPD ($OPD_m$) are in practice not recorded. As a result, information is missing to compute the shifted interferograms and to retrieve the corrected spectra from Equation (6). To do so, we would require accessing the interferograms at least until the maximal OPD plus the modulation frequency ($OPD_m + f$).

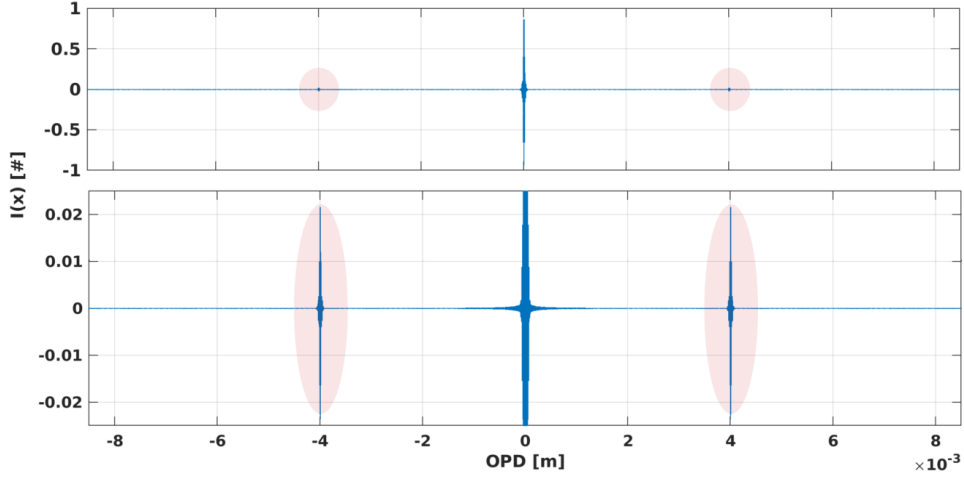

**Figure 1.** Example of simulated interferogram of a blackbody in presence of RTF modulation at frequency 4 mm and 5% relative amplitude. The bottom panel is identical to the upper one with the y-axis scaled to enhance the ghosts induced by the RTF modulation (underlined by the red ellipses)

This simple example shows that any correction should rely on retrieved signal frequencies above the instrument cut-off one, up to $OPD_m + f$ frequencies in this case. In other words, disentangling the RTF variations from the measured spectrum requires bringing higher frequency information about the input scene. A methodology to do so is proposed in the next section. Hereafter, we consider the general problem of any kind of RTF, i.e. not restricted to RTF modulations.

3.2. High-resolution estimate

The first step of the calibration ringing correction consists thus in estimating for any calibrated measurement $Sp_r(\nu)$ a high-resolution spectrum statistically representative of the input scene $Sp(\nu)$. This is achieved using the principal components (PC) decomposition of a dataset of high-resolution spectra $Sp_{ref}(\nu)$ spanning the range of viewing angles and of the scene diversity representative of the instrument to correct. This dataset can be generated using a RTM or can consist of measurements from other instruments with higher spectral resolution.

The principal components are defined as the eigenvectors associated to the greatest eigenvalues of the spectra covariance matrix [9]. The first PC represent the main directions of the signals, called the observation subspace, while the last ones only carry the noise [10]. Ten to a few hundreds PCs, noted $N_{PC}$, are generally sufficient to form an orthonormal basis of the high-resolution observation subspace. Noting $PC_{high,n}(\nu)$ the n$^{th}$ PC, it is then possible to compute the associated principal components at the instrument resolution and sampling $PC_{low,n}(\nu)$ by convoluting $PC_{high,n}(\nu)$ with the instrument SRF and applying a spline interpolation:

$$PC_{low,n}(\nu) = SPLINE\{[PC_{high,n} \otimes SRF](\nu)\} \quad (7)$$

Since the $PC_{low}$ are generally not an orthonormal basis, we re-normalize the $PC_{high}$ using the inverse of the $PC_{low}$ normalization matrix, noted $N$:

$$N(n,n') = \sum_\nu PC_{low,n}(\nu) \times PC_{low,n'}(\nu), \quad \{n,n'\} \in [1\ldots N_{PC}]^2$$

$$\widetilde{PC}_{high,n}(\nu) = \sum_{n'=1}^{N_{PC}} N^{-1}(n,n') \times PC_{high,n'}(\nu) \quad (8)$$

Every measurement is projected onto the basis $PC_{low}$ by computing the associated PC scores (PCS). The high-resolution statistical estimate $Sp_{guess}(v)$ can then be constructed using the same PCS but associated to the high-resolution basis $\widetilde{PC}_{high}$:

$$PCS(n) = \sum_v Sp_r(v) \times PC_{low,n}(v)$$
$$Sp_{guess}(v) = \sum_{n=1}^{N_{PC}} PCS(n) \times \widetilde{PC}_{high,n}(v) \qquad (9)$$

As a result, it is possible to estimate high-resolution spectra from lower resolution measurements bringing statistically relevant high frequency information. Nonetheless, the method is intrinsically limited: it is of course impossible to guess perfectly information that is not recorded by the instrument.

3.3. Correction factor

Assuming that the estimate $Sp_{guess}(v)$ is an adequate estimation of the input scene $Sp(v)$ and introducing a high-resolution reference of the instrument RTF $T_{ref}(v)$ representative of the actual RTF $T(v)$, we form the correction factor noted $\gamma(v)$ derived from Equation (2). It aims at cancelling the calibration and at retrieving a corrected spectrum $Sp_c(v)$ close to the input spectrum convoluted with the SRF.

$$\gamma(v) = \frac{[T_{ref} \otimes SRF](v) \times SPLINE\{[Sp_{guess} \otimes SRF](v)\}}{SPLINE\{[Sp_{guess}.T_{ref} \otimes SRF](v)\}}$$
$$Sp_c(v) = Sp_r(v) \times \gamma(v) = \frac{[Sp.T \otimes SRF](v)}{[T \otimes SRF](v)} \times \gamma(v) \cong [Sp \otimes SRF](v) \qquad (10)$$

The reference RTF $T_{ref}$ is not expected to vary rapidly and significantly in time. Moreover, it can be seen from Equation (10) that the correction factor is insensitive to low frequency RTF spectral variations in time, such as the presence of ice on the detectors would produce for example, since these variations would cancel-out. Thus, the reference transmission can be considered as quasi-static and would require only occasional updates. It can either be computed by oversampling an average of many calibration factors if the frequency of its spectral variations is lower than the maximum OPD of the instrument (which is the case for MTG-IRS for example), or come from instrument modelling or dedicated additional on-ground measurements otherwise.

In order to lighten the operational processing, the convolutions of the $\widetilde{PC}_{high}$ and $\widetilde{PC}_{high,n}.T_{ref}$ with the SRF can be pre-computed so that only the PCS for each measurement need to be calculated. The correction then writes as:

$$\gamma(v) = \frac{\sum_{n=1}^{N_{PC}} PCS(n) \times V_n(v)}{\sum_{n=1}^{N_{PC}} PCS(n) \times W_n(v)}$$
$$V_n(v) = [T_{ref} \otimes SRF](v) \times SPLINE\{[\widetilde{PC}_{high,n} \otimes SRF](v)\} \qquad (11)$$
$$W_n(v) = SPLINE\{[\widetilde{PC}_{high,n}.T_{ref} \otimes SRF](v)\}$$

With this approach, high-resolution guesses of the measurements can be estimated and used to construct a correction factor to mitigate the impact of the calibration ringing. It is fast enough to be used operationally as the computational burden imposed by the convolutions can be lightened using pre-computed coefficients. The methodology is referred to as *RTF uniformisation* as it mitigates the impact of the RTF for all spectral channels, as if the RTF was spectrally flat.

## 4. Simulations in the context of MTG-S IRS

In the following, we apply the RTF uniformisation to the case of MTG-S IRS LWIR band in order to demonstrate the ability and the efficiency of the method to mitigate the calibration ringing.

4.1. Simulation setup

The parameters of the simulation are such that the simulated instrument resembles to the MTG-S IRS LWIR band [3]. Nonetheless, the RTF used for the simulation is made-up specifically for this study and is not expected to be representative of the actual instrument. As shown on Figure 2, it consists of a smooth door multiplied by a modulation at frequency of 4 mm and a relative amplitude of 5% that is applied to all pixels. On Fig. 2, the instrument SRF has been plotted along with the RTF, clearly showing that the latter is varying at the scale of the SRF hence meaning that calibration ringing is expected.

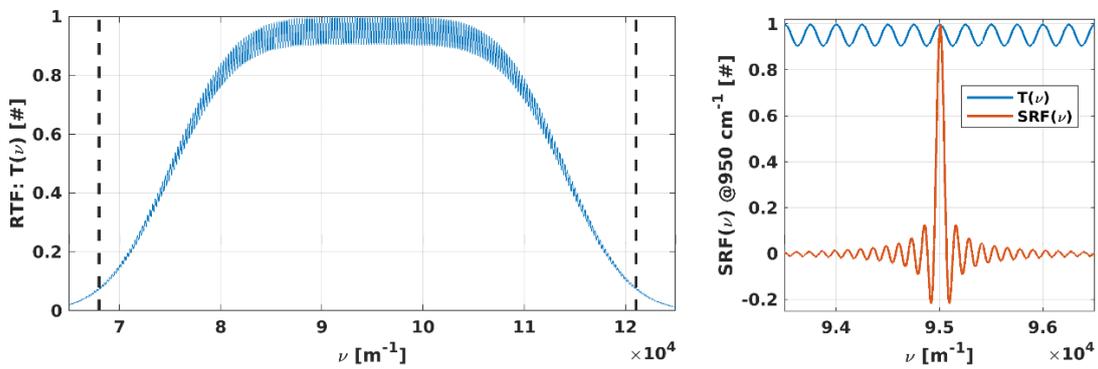

**Figure 2.** (**left panel**) Radiometric Transfer Function (RTF) as function of the wavenumber of the simulated instrument consisting in a smooth door multiplied by a modulation of 4 mm frequency and 5% relative amplitude. The dashed lines represent the MTG-S IRS wavenumber range accessible to the users. (**right panel**) Spectral Response Function (SRF) centered at 950 cm$^{-1}$ compared to the RTF.

The dataset to be corrected consists of a full Earth disc representative of what would be measured from the geostationary orbit at zero degree longitude by IRS. The Earth disc is covered with 280 step and stare dwells of 40x40 pixels, each pixel generating a spectrum. The number of pixels per dwell is down-scaled by a factor of 16 compared to IRS to ease the simulations. The input spectra are computed using a radiative transfer model (RTTOV [11]) set-up to generate IASI-like spectra between 650 and 1250 cm$^{-1}$ with a sampling of 0.25 cm$^{-1}$ for July the 14$^{th}$ 2022 at 12:00 UTC using ECMWF forecasts as inputs. First, the IASI numerical apodisation is removed at interferogram level. Then the simulated IASI spectra are multiplied by the RTF and applying a Fourier transform allows computing the associated raw interferograms. Computing MTG-S IRS-like spectra is then a matter of applying the IRS numerical apodisation and performing an inverse Fourier transform. Calibration views are generated along with the Earth views and are used for the radiometric calibration of the simulated products in a processing chain similar to the one that will be routinely operated [3]. Figure 3 shows the LWIR average spectral radiance distribution over the full disc and a selection of calibrated spectra sampled, like in reality, at 0.61 cm$^{-1}$. It is worth noting that the simulation is run without radiometric noise to emphasize the impact of the sole calibration ringing.

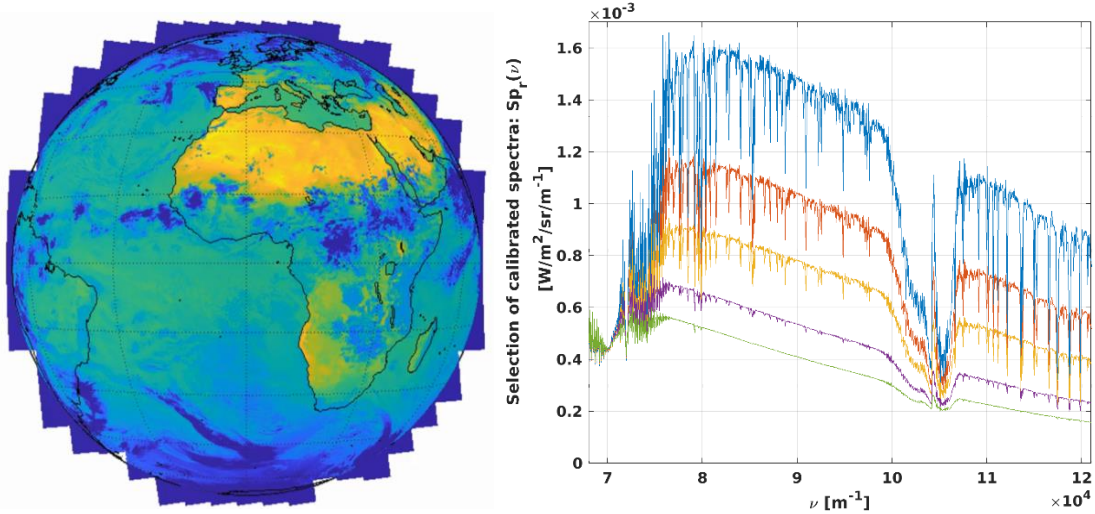

**Figure 3.** (**Left panel**) Average LWIR radiance of a full scan of MTG-S IRS consisting in 280 step and stare dwells. (**Right panel**) Selection of calibrated spectra over MTG-S IRS LWIR band representative of the simulation diversity.

4.2. RTF Uniformisation dataset

The RTF uniformisation dataset is composed of actual IASI measurements acquired from the Metop-C platform [1]. One full day of measurements on January the 1st 2023 was used, yielding approximately 1.200.000 spectra. IASI measurements are super-sampled enough with respect to IRS as IASI maximum OPD equals 20 mm which is significantly larger than the sum of the IRS maximum OPD (8.2 mm) and the transmission modulation frequency (4 mm), which is the criteria for an efficient correction as discussed in section 3.1. IASI spectra are then cropped to match the IRS LWIR band and the IASI numerical apodisation is removed.

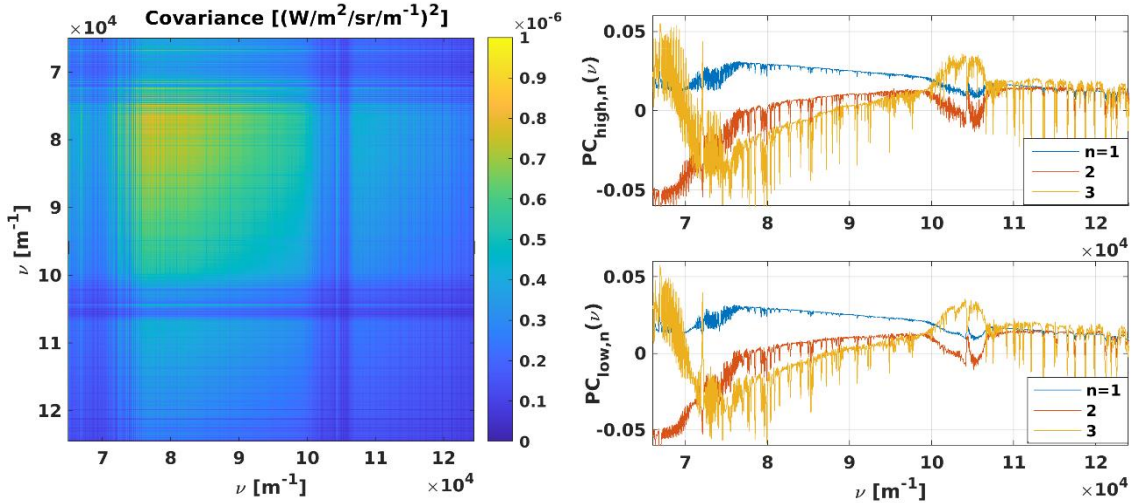

**Figure 4.** (**Left panel**) High-resolution covariance matrix over MTG-S IRS LWIR band. (**Right panel**) First three high and low resolution Principal Components (PC).

As described in section 3.2, we have computed the spectra covariance matrix and derived the $PC_{high}$ and associated $PC_{low}$. The high-resolution spectra covariance and the first three $PC_{high}$ and $PC_{low}$ are represented on Figure 4.

### 4.3. RTF Uniformisation efficiency

Choosing to construct the high-resolution dataset from actual IASI measurements and not from the same reference spectra used in the IRS simulation of section 4.1 avoids having perfect estimates by design and aims at proving that the efficiency of the RTF uniformisation is not overly sensitive to the choice of high-resolution dataset. Two simulations with and without the RTF uniformisation are run over the full IRS scan. The resulting calibrated spectra are then compared by subtracting them to a common reference free of calibration ringing. The derived radiometric errors are converted into equivalent temperature error dividing by the derivative of the Planck function at the temperature of reference of 280 K.

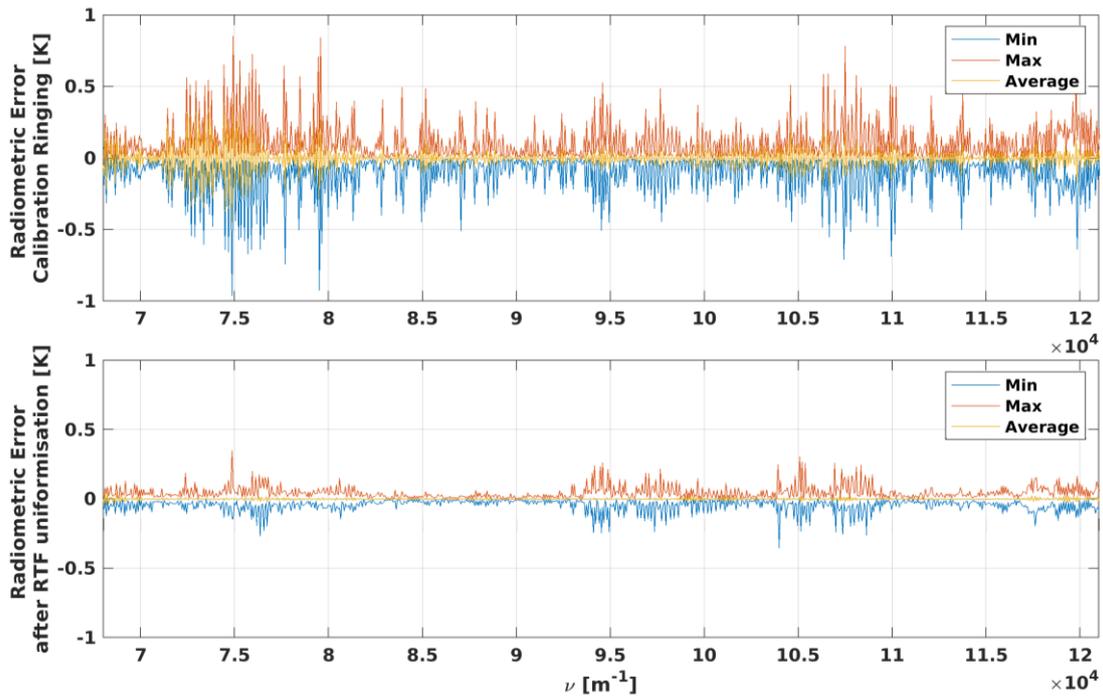

**Figure 5.** (**U**pper panel) Minimum, maximum and average radiometric error in equivalent temperature error in Kelvin due to the calibration ringing. (**B**ottom panel) Same metrics after application of the RTF uniformisation using 10 Principal Components (PC).

The minimum, maximum and average calibration ringing error as a function of the wavenumber is presented on Fig. 5. This highlights the strong scene dependency of the calibration ringing. The calibration ringing is characterized by error spikes up to $\pm 1$ K. After RTF uniformisation using 10 PC, a perfect cancellation of the average radiometric error and a strong reduction of the variability can be observed.

Fig. 6 shows the error standard deviation over the band computed from all spectra and expressed as equivalent temperature error as function of the number of PC. The calibration ringing is producing errors of 100 mK standard deviation. As soon as the RTF uniformisation is activated, this value rapidly decreases even if only a couple of PC are used, then reaches a plateau for which the error standard deviation is divided by a factor 10. It is worth noting that the error increases again when the number of PC reaches 50.

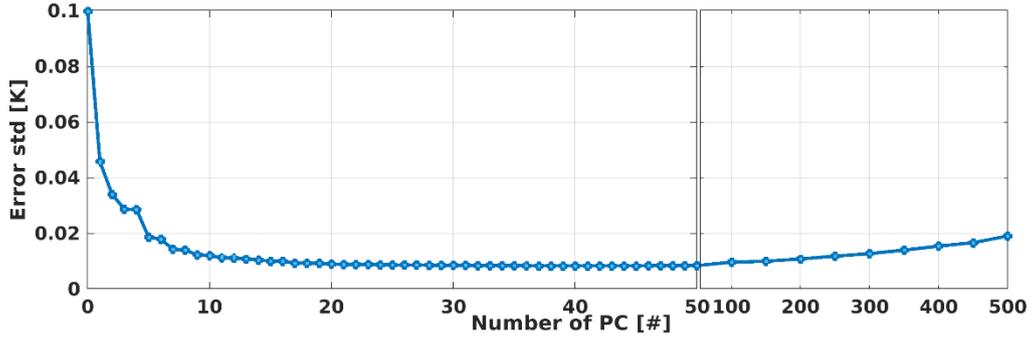

**Figure 6.** Residual ringing error standard deviation in Kelvin as function of the number of Principal Components (PC) used by the RTF uniformisation.

As discussed in section 3.2, the efficiency of the correction is naturally hampered by the intrinsic limitation in estimating the high-resolution spectrum. Nonetheless, the RTF uniformisation appears particularly efficient and well suited for an operational mitigation of the calibration ringing that could be observed in the MTG-S IRS LWIR band.

5. Discussion

There are still several points of limitations of this study and remarks for the future of calibration ringing mitigation in Fourier transform spectrometers:

- In view of the operational implementation of the RTF uniformisation by EUMETSAT for MTG-S IRS, the choice of the high-resolution dataset to use is open. The best candidates are Metop IASI [1], as presented in this study, which is considered as an international reference or its successor Metop-SG IASI-NG [12] (foreseen in 2025). The main limitation of these instruments are their maximum sounding angles (up to 50°) smaller than IRS (up to 90° close to earth rim); therefore, the dataset could be complemented with RTM simulations at high-sounding angles.

- The RTF uniformisation is based on the current knowledge of the instrument transmission. Thus, a careful monitoring of the calibration factors evolution in time is needed as well as updating the parameters of the algorithm if required. This technique would fail if for example the etalon characteristics would rapidly fluctuate in time, nonetheless that is not expected for MTG-S IRS.

- In real conditions, measurements are noisy; therefore, the high-resolution estimate can be adapted introducing a radiometric noise normalization into the PC projection of section 3.2. This point is not discussed in this study. Nonetheless, no particular impact on the efficiency of the RTF uniformisation is expected.

- The RTF uniformisation methodology is not specific to IRS LWIR band; it can be extended to its MWIR band and other hyperspectral instruments. It is also expected to be efficient to mitigate calibration ringing induced by band cut-off as for the CrIS instrument [6].

- The high-resolution statistical estimate approach introduced in section 3.2 is actually applicable to other hyperspectral instruments. It would help creating statistically relevant high-resolution datasets to test algorithms of new generation of satellites before launch.

6. Conclusions

In this paper, the fundamentals of calibration ringing, error have been recalled: a radiometric error propagated to the calibrated radiances acquired by FTS caused by significant spectral variations of the instrument RTF at the scale of the instrument SRF. A mitigation strategy, called RTF uniformisation, relying on a high spectral resolution estimate of the measurements using principal components decomposition, has been introduced. Finally, the RTF uniformisation efficiency has been assessed in the context of MTG-S IRS LWIR band.

The RTF uniformisation appears sufficiently efficient and computationally inexpensive to be implemented into the operational processing of MTG-S IRS as a post-processing applied to the LWIR calibrated radiances. Doing so, a reduction of the radiometric errors induced by calibration ringing by a factor 10 can be expected.

**Author Contributions:** Conceptualization, B.To., C.S. and P.D.; investigation and writing P.D.; coding, P.D. and G.D.; review and editing, G.D., D.C.,C.S. and B.Th.; project administration, D.C.; All authors have read and agreed to the published version of the manuscript.

**Funding:** This research was funded by EUMETSAT.

**Data Availability Statement:** Data used in this study are described in section 4 and accessible on EUMETSAT data center (https://www.eumetsat.int/eumetsat-data-centre).

**Acknowledgments:** We would like to thank Dave Tobin (CIMSS) and Nigel Atkinson (Met Office) for initial discussion on this topic.

**Conflicts of Interest:** The authors declare no conflict of interest.

**References**

1. Blumstein, D., Chalon, G., Carlier, T., Buil, C., Hebert, P., Maciaszek, T., ... & Jegou, R. (2004). IASI instrument: Technical overview and measured performances. Infrared Spaceborne Remote Sensing XII, 5543, 196-207.
2. Goldberg, M. D., Kilcoyne, H., Cikanek, H., & Mehta, A. (2013). Joint Polar Satellite System: The United States next generation civilian polar-orbiting environmental satellite system. Journal of Geophysical Research: Atmospheres, 118(24), 13-463.
3. MTG-S IRS Level-1 Algorithm Theoretical Basis Document: https://www.eumetsat.int/media/50577
4. Dussarrat, P., Theodore, B., Coppens, D., Standfuss, C., & Tournier, B. (2021). Introduction to the calibration ringing effect in satellite hyperspectral atmospheric spectrometry. arXiv preprint arXiv:2111.08574.
5. Lee, L., Qi, C., & Ding, L. (2023, January). The instrumental responsivity effect to the calibrated radiances of infrared hyperspectral benchmark sounder. In Earth and Space: From Infrared to Terahertz (ESIT 2022) (Vol. 12505, pp. 645-653). SPIE.
6. Borg, L., Loveless, M., Knuteson, R., Revercomb, H., Taylor, J., Chen, Y., ... & Tobin, D. (2023). Simulation of CrIS Radiances Accounting for Realistic Properties of the Instrument Responsivity That Result in Spectral Ringing Features. Remote Sensing, 15(2), 334.
7. Wikipedia webpage of Fabry-Perot and etalon: https://en.wikipedia.org/wiki/Fabry%E2%80%93P%C3%A9rot_interferometer
8. Coppens, D., Theodore, B., & Klaes, D. (2017, September). MTG-IRS: from raw measurements to calibrated radiances. In Earth Observing Systems XXII (Vol. 10402, pp. 12-19). SPIE.
9. Wikipedia webpage of Principal component analysis: https://en.wikipedia.org/wiki/Principal_component_analysis
10. Antonelli, P., Revercomb, H. E., Sromovsky, L. A., Smith, W. L., Knuteson, R. O., Tobin, D. C., ... & Best, F. A. (2004). A principal component noise filter for high spectral resolution infrared measurements. Journal of Geophysical Research: Atmospheres, 109(D23).
11. Saunders, R., Hocking, J., Turner, E., Rayer, P., Rundle, D., Brunel, P., ... & Lupu, C. (2018). An update on the RTTOV fast radiative transfer model (currently at version 12). Geoscientific Model Development, 11(7), 2717-2737.
12. Crevoisier, C., Clerbaux, C., Guidard, V., Phulpin, T., Armante, R., Barret, B., ... & Stubenrauch, C. (2014). Towards IASI-New Generation (IASI-NG): impact of improved spectral resolution and radiometric noise on the retrieval of thermodynamic, chemistry and climate variables. Atmospheric Measurement Techniques, 7(12), 4367-4385.